# FundaQ-8: A Clinically-Inspired Scoring Framework for Automated Fundus Image Quality Assessment

Lee Qi Zun
*DR.MATA Team*
*Qmed Asia*
*Universiti Utara Malaysia*
Kuala Lumpur, Malaysia
Email: jacky@qmed.asia

Dr. Zalifa Zakiah Binti Asnir
*DR.MATA Team (Consultant Opthalmologist)*
*Hospital Cyberjaya*
Selangor, Malaysia
Email: zalifazakiah@moh.gov.my

Oscar Wong Jin Hao
*DR.MATA Team*
*Qmed Asia*
Kuala Lumpur, Malaysia
Email: oscar@qmed.asia

Dr. Mohamad Sabri bin Sinal @ Zainal
*Senior Lecturer*
*Universiti Utara Malaysia*
Kedah, Malaysia
Email: msabri@uum.edu.my

Dr. Nor Anita Binti Che Omar
*DR.MATA Team (Consultant Opthalmologist)*
*Hospital Sultanah Nur Zahirah*
Kuala Terengganu, Malaysia
Email: drnoranita@moh.gov.my

Goh Man Fye
*DR.MATA Team*
*Qmed Asia*
Kuala Lumpur, Malaysia
Email: manfye@qmed.asia

*Abstract*— Automated fundus image quality assessment (FIQA) remains challenging due to variations in image acquisition and subjective expert evaluations. This study introduces FundaQ-8, a novel, expert-validated framework designed to systematically assess fundus image quality based on eight critical parameters, including field coverage, anatomical visibility, and others. Using FundaQ-8 as a structured scoring reference, a ResNet18-based regression model is built to predict continuous quality scores (0–1), trained on 1,800 fundus images comprising real-world clinical Kaggle data. The model leverages transfer learning with an MSE-optimized regression adaptation, and standardized preprocessing. Validation against the EyeQ dataset and statistical analyses confirm the framework's reliability in providing interpretable, clinically meaningful assessments. Furthermore, our findings indicate that incorporating FundaQ-8 as the scoring framework for FIQA into deep learning-based diabetic retinopathy (DR) grading models enhances diagnostic robustness.

## I. Introduction

Ocular diseases such as glaucoma, diabetic retinopathy (DR), and age-related macular degeneration (AMD) are leading causes of irreversible vision loss, affecting over 200 million people globally [1]. Studies indicate that nearly half of participants in screening programs have at least one eye disorder, primarily refractive errors [2]. Retinal imaging plays a critical role in early detection, with color fundus photography (CFP) being the most widely used modality due to its cost-effectiveness and ability to reveal pathological features. While advanced imaging techniques like OCT and wide-angle imaging offer greater detail, they come with drawbacks such that OCT is costly and less portable, while wide-angle imaging compromises resolution [3]. Consequently, CFP remains the primary tool for large-scale screenings.

However, the diagnostic utility of fundus imaging is heavily dependent on image quality, which is often degraded by motion artifacts, improper focus, uneven illumination, and obstructions like cataracts [4][5]. Studies show that 12–25% of fundus images in screening programs fail to meet diagnostic quality standards, necessitating retakes that delay diagnosis and increase costs [5][6].

Despite advancements in camera technology, image quality remains inconsistent due to variability in operator skill, patient cooperation, and environmental conditions.

Automated fundus image quality assessment (FIQA) aims to address these challenges by providing real-time, objective feedback on image usability. Traditional rule-based methods, which rely on handcrafted features such as illumination and contrast, often struggle to generalize across different populations and camera systems [6][7]. Manual grading by ophthalmologists, though clinically validated, is labor-intensive and subjective, with studies reporting significant inter-rater variability in quality assessments [6].

FIQA is particularly crucial in the context of automated DR screening, where grading systems have significantly improved efficiency and scalability [14]. However, the reliability of these systems is highly dependent on input image quality. Low-quality images can lead to misdiagnoses—false positives may cause unnecessary anxiety, while false negatives can delay critical treatment. Integrating FIQA ensures that only diagnostically viable images are analyzed, enhancing the accuracy of DR detection and reducing diagnostic errors.

To address these challenges, we propose FundaQ-8 (Fundus Quality 8-Attribute Scoring), a novel FIQA framework that employs an eight-point quality scoring system for fundus images. Built on a ResNet18 architecture with transfer learning in a regression framework [8], FundaQ-8 leverages a multi-source dataset with expert-validated quality labels to ensure objective assessment. By bridging the gap between clinical requirements and computational solutions, FundaQ-8 enhances image quality evaluation, ultimately improving diagnostic accuracy and workflow efficiency

## II. Related Work

### A. CNN Architectures for FIQA

Recent studies on automated fundus image quality assessment (FIQA) have leveraged both traditional machine learning (ML) and deep learning (DL) techniques, with convolutional neural networks (CNNs) emerging as the most effective tools due to their ability to learn hierarchical features directly from raw images [9]. For example, ensembles of EfficientNetV2 models have demonstrated superior accuracy compared to single models on challenging datasets such as DeepDRiD [10]. Similarly, ResNet architectures, particularly a fine-tuned ResNet50 achieving 98.6% accuracy, have proven adept at mitigating issues like the vanishing gradient problem and capturing subtle degradations such as blur and uneven illumination that affect image quality [10]. However, while these architectures offer robust feature extraction, their performance often depends on dataset characteristics and specific task requirements.

### B. FIQA Algorithms and Grading Scales

Complementing these architectural advances, FIQA algorithms have increasingly integrated subjective image quality assessments to generate expert-validated ground truth labels for model training. Key perceptual parameters such as anatomical visibility, sharpness, color fidelity, and contrast are incorporated through weighted scoring methods to better reflect clinical evaluations [9]. Some approaches further enhance clinical relevance by mimicking the human visual system (HVS) with features like multi-channel sensation, just-noticeable blur, and contrast sensitivity functions [11]. Moreover, grading scales ranging from continuous 0.0 - 1.0 metrics to multi-attribute systems (evaluating focus, illumination, field definition, and artifacts) have been developed to standardize quality assessments [12]. Nevertheless, many existing systems remain limited by subjectivity and inter-rater variability, with datasets often biased toward specific imaging conditions [13].

### C. Limitations of Current FIQA Methods

Despite these advancements, current FIQA methods still face limitations. For instance, subjective quality assessment by experts is time-consuming and can introduce variability, while many datasets are biased toward specific imaging conditions, limiting the generalizability of existing models [13]. Additionally, generic feature-based methods may not capture the nuanced structural details essential for diagnosis and are often sensitive to noise. These underscore a critical research gap which is the need for a robust, objective, and scalable FIQA framework that minimizes subjectivity and adapts to diverse imaging environments. Our study addresses this gap by introducing FundaQ-8, a novel eight-attribute scoring framework that provides continuous, interpretable quality scores and supports enhanced automated diabetic retinopathy (DR) grading by filtering out low-quality images.

## III. METHODOLOGY

This section describes the proposed FundaQ-8 framework for automated fundus iamge quality assessment, detailing the dataset, preprocessing steps, scoring mechanism, and evaluation criteria. The detailed procedure is shown in the subsections below.

### A. Dataset Collection and Preprocessing

The dataset for this study comprises 1,800 fundus images sourced from two distinct origins to ensure diversity. A total of 1,393 anonymized patient images were collected from local hospital under the supervision of experienced ophthalmologists, capturing real-world variations in quality. These images were acquired using the Topcon NW8F fundus camera, a widely used device to capture detailed retinal images. To complement this, 407 standardized images were obtained from the publicly available Standardized Multi-Channel Dataset for Glaucoma (SMDG-19) on Kaggle, which provides preprocessed images acquired under controlled conditions [15].

Local images then underwent a three-step preprocessing pipeline to standardize inputs: (1) black region cropping to isolate the retina, (2) aspect ratio correction to a 1:1 square format, and (3) resizing to 512×512 pixels for compatibility with deep learning architectures. Kaggle images, already standardized for field-of-view, required only minimal adjustments. This approach ensured the model's exposure to both clinical variability and controlled quality benchmarks.

### B. FundaQ-8 Scoring Framework

To systematically evaluate fundus image quality, this study introduces FundaQ-8, a scoring framework developed in collaboration with ophthalmologists from our team with over 30 years of clinical experience. FundaQ-8 assesses images based on eight critical quality parameters. Parameters were scored on a 0–2 Likert scale (e.g., 2=optimal, 0=unusable), with criteria explicitly tied to diagnostic requirements as shown in TABLE I. Total scores were normalized to a continuous 0–1 scale, where 1.0 represents perfect quality. The scoring process included iterative reviews to ensure inter-rater reliability.

TABLE I. FUNDAQ-8 SCORING FRAMEWORK FOR FUNDUS IMAGE QUALITY ASSESSMENT

| Field | Score Range | Scoring Description |
|---|---|---|
| 1. Resolution (Blurry) | 0 - 2 | 2: Very clear, sharp details visible. |
| | | 1: Clear but slightly blurry; details still distinguishable. |
| | | 0: Blurry; key features difficult to distinguish. |
| 2. Field of View (Coverage) | 0 – 2 | 2: Full coverage, including optic disc, macula, and peripheral retina. |
| | | 1: Partial coverage; optic disc and macula visible, but periphery incomplete. |
| | | 0: Incomplete; key features missing (e.g., macula not visible). |
| 3. Color Fidelity | 0 – 2 | 2: Natural, realistic colors; no discoloration. |
| | | 1: Slight discoloration; features still identifiable. |
| | | 0: Severe discoloration; features difficult to identify. |
| 4. Presence of Artifacts | 0 - 2 | 2: No artifacts; image is clean. |
| | | 1: Minor artifacts (e.g., slight glare or dust) that do not obscure key features. |
| | | 0: Major artifacts obscure key features (e.g., glare or smudges). |
| 5. Vessels | 0 – 2 | 2: Vessels are clearly visible with good contrast. |
| | | 1: Vessels are partially visible; slight loss of detail. |
| | | 0: Vessels are blurry or indistinguishable. |
| 6.. Macula | 0 – 2 | 2: Macula is clearly visible with defined edges. |
| | | 1: Macula is visible, but slightly blurred or with some detail loss. |
| | | 0: Macula is not visible or indistinct. |
| 7. Optic Disc | 0 – 2 | 2: Optic disc is clearly visible with distinct edges. |
| | | 1: Optic disc is visible, but slightly blurred or with some detail loss. |
| | | 0: Optic disc is not visible or indistinct. |
| 8. Optic Cup | 0 - 2 | 2: Optic cup is clearly visible and distinguishable from the disc. |

| | | 1: Optic cup is visible, but less defined. |
| | | 0: Optic cup is not visible or indistinct. |

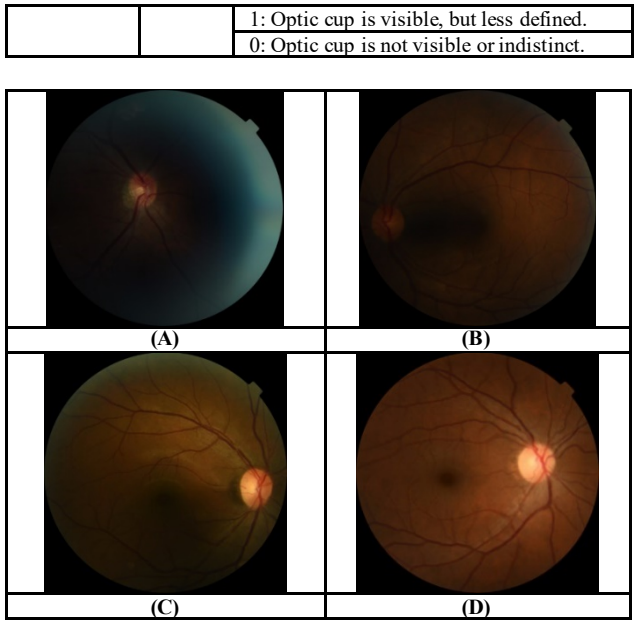

Fig. 1. Example of fundus images **(A), (B), (C), (D)** with different quality from collected dataset.

TABLE II.    IMAGE QUALITY SCORE FOR FUNDUS IMAGES (A), (B), (C), (D) BASED ON THE PROPOSED FUNDAQ-8 SCORING FRAMEWORK.

| Image | 1 | 2 | 3 | 4 | 5 | 6 | 7 | 8 | Total /16 | Score (0-1) |
|---|---|---|---|---|---|---|---|---|---|---|
| (A) | 0 | 0 | 0 | 0 | 0 | 0 | 2 | 2 | 4 | 0.25 |
| (B) | 1 | 2 | 0 | 1 | 2 | 0 | 1 | 1 | 8 | 0.5 |
| (C) | 2 | 2 | 0 | 2 | 2 | 0 | 2 | 2 | 12 | 0.75 |
| (D) | 2 | 2 | 2 | 2 | 2 | 2 | 2 | 2 | 16 | 1 |

### C. Model Architecture and Adaptation

To enhance the accuracy and automation of fundus image quality assessment, a deep learning-based approach was adopted. The ResNet18 architecture (as shown in Fig. 2) was selected for its proven efficacy in medical imaging tasks and computational efficiency. Pre-trained weights from ImageNet were retained to leverage learned feature extraction capabilities, while the final fully connected layer was replaced with a regression head (single neuron output) to predict continuous quality scores. Transfer learning was prioritized over training from scratch to mitigate overfitting risks given the moderate dataset size.

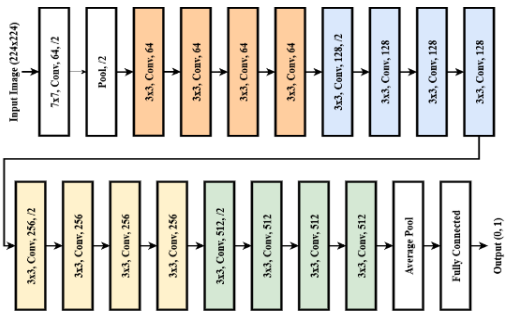

Fig. 2. ResNet18 Architecture with FC Layer for Regression Task.

### D. Dataset Partitioning and Model Training

To develop and evaluate the FundaQ-8 scoring model, the dataset was partitioned into training, validation, and test sets. A stratified sampling approach was applied to maintain a balanced distribution of quality scores across all subsets, ensuring that the model generalizes well across different quality levels. The dataset was partitioned into training (70%, 1,260 images), validation (15%, 270 images), and test (15%, 270 images) sets, with stratification to maintain consistent score distributions across splits. Training employed the Adam optimizer with a learning rate of $1\times10^{-4}$ and weight decay of $1\times10^{-5}$ to regularize weights. A batch size of 4 was used to balance GPU memory constraints. The model was trained for 50 epochs using MSE loss. Training was halted if no improvement occurred for 10 consecutive epochs. The distributions of scores for the dataset are shown in Fig. 3.

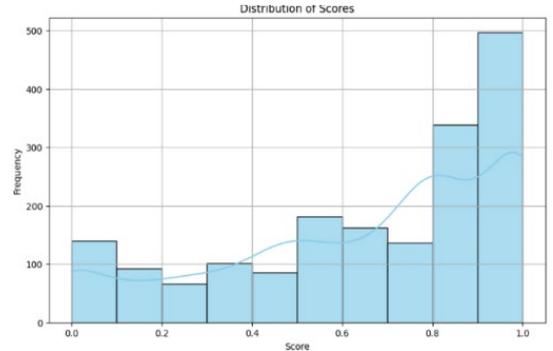

Fig. 3. Histogram detailing distribution of scores for datasets. (n=1,800)

### E. Model Evaluation

The model will be evaluated using a three-pronged approach: first, through quantitative evaluation metrics; second, by comparing the predicted quality scores with the Eye-Quality Assessment Dataset; and third, by analyzing its impact on automated diabetic retinopathy detection.

*a) Quantitative Evaluation Metrics*

The predictive accuracy was assessed using several quantitative evaluation metrics which included the Mean Squared Error (MSE), Mean Absolute Error (MAE), Root Mean Squared Error (RMSE), $R^2$ (Coefficient of Determination) and Explained Variance score. These metrics offered a comprehensive understanding of the model's ability to predict the quality of fundus images accurately, in alignment with expert assessments.

*b) Comparison with Eye-Quality Categorical Data*

Eye-Quality (EyeQ) Assessment Dataset, is a re-annotated subset of EyePACS comprises 28,792 retinal images categorized into three quality levels: 0 ('Good'), 1 ('Usable'), and 2 ('Rejected') [9]. The model's continuous predicted quality scores were then compared with these discrete EyeQ categories. Statistical analyses were conducted to assess the relationship between the predicted scores and EyeQ labels, including Spearman correlation to measure association strength, ANOVA to test for significant differences across categories, and Ordinary Least Squares (OLS) regression to quantify the contribution of EyeQ categories to predicted scores.

*c) Impact on Automated Diabetic Retinopathy Detection*

The EyeQ dataset contained diabetic retinopathy (DR) grades ranging from 0 to 4, which were simplified into three classes: grade 0 remained unchanged, grades 1 and 2 were merged into grade 1, and grades 3 and 4 into grade 2. Then, a deep learning-based DR detection model was applied to the dataset. Images were categorized into three quality groups based on the predicted quality score: Bad (score<0.4), Medium (0.4<score<0.8), and Good (≥0.8). This approach provide analysis on how image quality variations influence DR detection performance.

To determine the extent to which image quality affects automated DR detection outcomes, evaluation metrics such as accuracy, sensitivity, specificity, and F1-score, were analyzed for each quality group. Accuracy measured classification performance, sensitivity assessed the model's ability to detect DR cases, specificity evaluated its ability to exclude non-DR cases, and the F1-score provided a balanced assessment of precision and recall.

## IV. TECHNICAL IMPLEMENTATION & EXPERIMENT

### A. Softwares and Libraries

The framework was implemented in Python 3.10 using PyTorch 2.2.2 with CUDA 12.1 for GPU acceleration to optimize tensor operations. Data preprocessing and analysis utilized pandas (v2.2.3) for dataset structuring, while visualization libraries (matplotlib v3.9.2 and seaborn v0.13.2) generated score distributions and training curves. Performance metrics (MAE, MSE, $R^2$) were computed using scikit-learn (v1.5.2). Model weights were initialized from PyTorch's pretrained ResNet18 implementation in torchvision (v0.17.2).

### B. Hardware Configuration

All experiments were conducted on a laptop equipped with an AMD Ryzen 9 5900HX processor (8 cores, 16 threads, 3.3–4.6 GHz). The system featured an NVIDIA GeForce RTX 3060 Laptop GPU with 6GB GDDR6 VRAM and 3,840 CUDA cores. Additionally, it was equipped with 16GB of DDR4 RAM (3,200 MHz) and a 1TB NVMe SSD.

## V. RESULTS

This section presents the evaluation results of the proposed FundaQ-8 framework, focusing on its performance in fundus image quality assessment. The analysis includes a quantitative evaluation of the model's predictive accuracy, a comparison between the model-predicted quality scores and the EyeQ dataset's quality categories, and an assessment of its impact on automated diabetic retinopathy detection. The detailed results are discussed in the subsections below.

### A. Quantitative Evaluation of the Model

The proposed ResNet18-based model achieved robust performance on the test set, demonstrating strong alignment with expert quality assessments. Quantitative evaluation yielded the following metrics as shown in Table III.

TABLE III. EVALUATION METRICS ON TEST SET (N=270)

| Metrics | Value |
|---|---|
| Mean Squared Error (MSE) | 0.0217 |
| Mean Absolute Error (MAE) | 0.0992 |
| Root MSE (RMSE) | 0.1473 |
| R² (Coefficient of Determination) | 0.7734 |
| Explained Variance | 0.7737 |

The model's performance was rigorously evaluated using regression-specific metrics to quantify its alignment with expert quality assessments (Table III). The MSE of 0.0217 indicates minimal deviation, with outliers penalized quadratically to prevent severe misgradings. The MAE of 0.0992 (9.92% error) falls below the 15% threshold considered clinically acceptable, while the RMSE of 0.1473 confirms stable performance. An $R^2$ score of 0.7734 and an Explained Variance of 0.7737 validate the model's ability to explain 77.34% of variance in expert ratings with minimal bias. Additionally, a median absolute error of 0.0631 highlights the model's precision in most test cases.

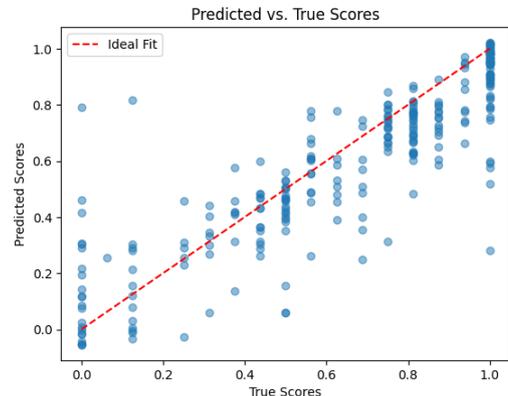

Fig. 4. Predicted vs. True Scores for test set, (n=270). It reveals tight clustering around the y=x line when tested on the test set.

### B. Comparison of EyeQ Quality Categories and Model Predicted Quality Score

In this section, the relationship between the EyeQ quality categories and the predicted Quality Score from the proposed model is analyzed. The EyeQ quality categories consist of three distinct levels (0, 1, and 2) while the predicted Quality Score is a continuous variable produced by the proposed model, with smaller values signifying poorer quality and larger values indicating better quality. Table IV displays the statistical analysis of the relationship.

TABLE IV. STATISTICAL ANALYSIS OF PREDICTED QUALITY SCORES AND EYEQ CATEGORIES

| Statistical Analysis | Metrics | Value |
|---|---|---|
| Spearman Correlation | Coefficient | -0.752 |
| ANOVA Test | F-statistic | 28,380.910 |
|  | p-value | 0.000 |
| OLS Regression | R² | 0.653 |
|  | Coefficient | -0.265 |
|  | p-value | 0.000 |

The Spearman correlation coefficient of -0.752 indicates a strong negative relationship and demonstrates the model's ability to capture the inverse relationship between image quality and the EyeQ categories. An ANOVA test confirmed the statistical significance of the differences in predicted scores across categories, with an F-statistic of 28,380.910 and a p-value of 0.0, indicating a meaningful variation between groups. The OLS regression analysis showed that the EyeQ categories explained 65.3% of the variance in the predicted scores ($R^2 = 0.653$). The regression coefficient of -0.265 indicates that each increase in the EyeQ category corresponds to an average decrease of 0.2653 units in the predicted score, with a statistically significant p-value of 0.000.

The box plot as shown in Fig.5 illustrates the relationship between EyeQ categories and predicted Quality Scores. Category 0, representing the highest quality images, has higher and more consistent predicted scores, indicating reliable performance for high-quality inputs. Conversely, category 2, representing the lowest quality images, has lower predicted scores and greater variability, reflecting challenges in predicting poorer-quality inputs.

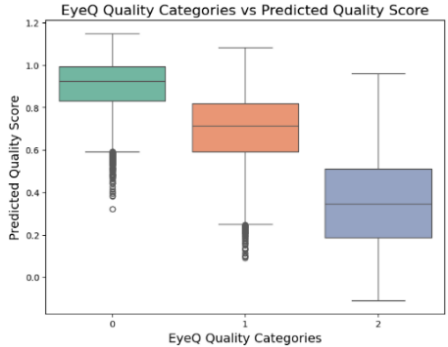

Fig. 5. Box plot of EyeQ Quality Categories vs. Predicted Quality Scores by Proposed Model. (n=28,792)

Notably, there are several outliers in category 0 (Good) and category 1 (Usable), which suggests some discrepancies between the manually labeled EyeQ quality categories and the predicted quality scores. Several examples of outliers can be seen in Fig. 6. The proposed framework in this study labels images based on an eight-attribute scoring system, ensuring a more objective assessment of image quality. In contrast, the EyeQ dataset was manually labeled by ophthalmologists, introducing potential subjectivity in the quality classification process. This difference in approach explains some of the outliers observed in the analysis.

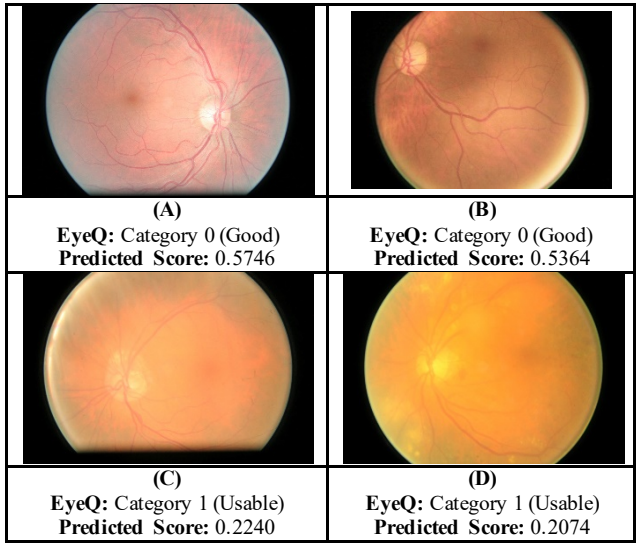

Fig. 6. Examples of outliers in boxplot with predicted score by proposed model for labelled fundus images in EyeQ dataset. Images **(A)**, **(B)** are labelled as category 0 while images **(C)**, **(D)** falls in category 1.

### C. Impact on Automated Diabetic Retinopathy Detection

The performance of the DR detection model (Table V) was evaluated within each quality group by comparing the predicted DR grades to the ground truth DR grades.

TABLE V. PERFORMANCE METRICS OF DIABETIC RETINOPATHY DETECTION ON DIFFERENT QUALITY GROUPS (N=28,792)

| Quality Group | Accuracy | Sensitivity | Specificity | F1-Score |
|---|---|---|---|---|
| Bad (Score < 0.4) | 0.5464 | 0.5464 | 0.5464 | 0.5464 |
| Medium (0.4 < Score < 0.8) | 0.4846 | 0.4846 | 0.4846 | 0.4846 |
| Good (Score > 0.8) | 0.7357 | 0.7357 | 0.7357 | 0.7357 |

These results indicated that image quality plays a significant role in the performance of automated DR detection. Good quality images yielded the highest metrics, with an accuracy of 0.7832, sensitivity of 0.6108, specificity of 0.8136, and F1-score of 0.6258, suggesting that clearer images enhance DR detection. Conversely, Bad quality images significantly reduced performance, with an accuracy of 0.5464, sensitivity of 0.4846, and F1-score of 0.4152, highlighting the difficulty in accurately identifying DR in these images. Medium quality images exhibited intermediate performance, further supporting the trend of improved DR detection with increasing image quality.

## VI. DISCUSSIONS

The proposed ResNet18-based model, integrated with the FundaQ-8 scoring system, demonstrates the feasibility of automated fundus image quality assessment (FIQA) while maintaining strong alignment with expert evaluations. By leveraging transfer learning and regression-based scoring, FundaQ-8 provides a more granular and interpretable quality assessment compared to traditional categorical grading. The model achieved a MAE of 0.0992 (9.92% error), well within the clinically acceptable 15% threshold, with an $R^2$ score of 0.7734, capturing over 77% of the variance in expert-assigned scores. The low MSE (0.0217) and RMSE (0.1473) indicate minimal large errors that could impact clinical decision-making, reinforcing the model's reliability for real-world applications.

Beyond numerical performance, FundaQ-8 enhances clinical applicability by ensuring consistency in quality assessment across varied datasets. Validation against the categorical EyeQ dataset showed a strong negative Spearman correlation (-0.7524) with quality categories, confirming its effectiveness in distinguishing image quality levels. Statistical analyses, including ANOVA and OLS regression, further validated its robustness, with EyeQ categories accounting for 65.3% of the variance in predicted scores. This suggests that FundaQ-8 can effectively standardize quality evaluation, minimizing subjectivity in expert grading and facilitating large-scale retinal imaging studies.

In the context of automated diabetic retinopathy (DR) detection, FundaQ-8 plays a crucial role in optimizing diagnostic accuracy. Poor-quality images increase misclassification risks, leading to unnecessary retakes and delayed treatment. Our analysis reveals that high-quality images yield a DR detection accuracy of 78.3%, while medium- and low-quality images drop to 63.4% and 54.6%, respectively. By filtering out low-quality inputs, FundaQ-8 not only enhances model performance but also improves resource efficiency in screening programs. Its integration into automated DR grading workflows ensures that only diagnostically useful images contribute to decision-making, ultimately reducing false positives and false negatives while supporting timely patient care. Thus, FundaQ-8 represents a significant advancement in FIQA, providing an interpretable, scalable, and clinically relevant solution for improving retinal imaging quality assessment.

## VII. Limitations and Future Works

While the proposed framework shows promise, its clinical applicability is limited by dataset diversity, lacking rare pathologies and extreme degradations from factors like cataracts and device artifacts. These cases pose challenges the model may not have learned to recognize, impacting generalizability. Additionally, potential overfitting to specific imaging characteristics may reduce adaptability across varied clinical settings. Future work should expand dataset diversity, include rare cases, and develop domain adaptation techniques to improve robustness in real-world environments.

## VIII. Conclusion

This study introduced FundaQ-8, a novel scoring framework for automated fundus image quality assessment (FIQA) that standardizes evaluation while reducing reliance on expert annotations. Unlike traditional categorical grading, FundaQ-8 assigns a continuous 0–1 quality score based on eight key attributes, enabling precise and objective assessment. The ResNet18-based model, trained using FundaQ-8 labels, demonstrated strong alignment with expert evaluations, achieving a low MAE and a robust $R^2$ score. Validation against the EyeQ dataset further confirmed its reliability. Given that 12–25% of fundus images in screening programs require retakes, delaying diagnosis and increasing costs, FundaQ-8 helps mitigate inefficiencies by accurately detecting low-quality images and optimizing clinical workflows. Beyond FIQA, its integration into automated diabetic retinopathy (DR) screening enhances DR detection accuracy and reduces unnecessary re-screening, ultimately improving early diagnosis and patient outcomes. In conclusion, FundaQ-8 represents a significant advancement in FIQA, offering a reliable and interpretable solution to improve retinal imaging in clinical settings.

## Acknowledgment

The research team sincerely appreciates the invaluable contributions of the ophthalmologists from DR.MATA team who played a crucial role in designing the scoring framework and providing expert annotations for the dataset. Their expertise and dedication were instrumental in ensuring the accuracy and clinical relevance of this study. We also extend our gratitude to all team members whose efforts in model development, data processing, and analysis made this work possible.